\documentclass[pra,twocolumn,showpacs,amsmath]{revtex4}
\usepackage{graphicx}

\def\nn{\nonumber}
\def\l{\left}
\def\r{\right}
\newcommand{\bra}[1]{\langle #1 | \,}
\newcommand{\ket}[1]{\, | #1 \rangle}

\newcommand{\expv}[1]{\langle #1 \rangle}
\newcommand{\ve}{\epsilon}

\newcommand{\tJ}{\tilde{J}}
\newcommand{\hlf}{\mbox{$\frac{1}{2}$}}
\newcommand{\sgn}{\operatorname{sgn}}
\newcommand{\arccosh}{\operatorname{arccosh}} 
\newcommand{\us}{\uparrow}
\newcommand{\ds}{\downarrow}
\newcommand{\be}{\begin{equation}}
\newcommand{\ee}{\end{equation}}
\newcommand{\bea}{\begin{eqnarray}}
\newcommand{\eea}{\end{eqnarray}}
\newcommand{\beann}{\begin{eqnarray*}}
\newcommand{\eeann}{\end{eqnarray*}}
\newcommand{\besal}[1]{\begin{subequations}\label{#1}\begin{eqnarray}}
\newcommand{\besa}{\begin{subequations}\begin{eqnarray}}
\newcommand{\eesa}{\end{eqnarray}\end{subequations}}

\begin{document}

\title{Attractively bound pairs of atoms in the Bose-Hubbard model 
and antiferromagnetism}

\author{Bernd Schmidt}
\author{Michael Bortz}
\author{Sebastian Eggert}
\author{Michael Fleischhauer}
\affiliation{Fachbereich Physik, Technische Universit\"at Kaiserslautern, 
D-67663 Kaiserslautern, Germany}
\author{David Petrosyan} 
\affiliation{Institute of Electronic Structure \& Laser, FORTH, 
71110 Heraklion, Crete, Greece}

\date{\today}

\begin{abstract}
We consider a periodic lattice loaded with pairs of bosonic atoms
tightly bound to each other via strong attractive on-site interaction 
that exceeds the inter-site tunneling rate. An ensemble of such 
lattice-dimers is accurately described by an effective Hamiltonian 
of hard core bosons with strong nearest-neighbor repulsion which is
equivalent to the $XXZ$ model with Ising-like anisotropy. We calculate 
the ground-state phase diagram for a one-dimensional system which 
exhibits incompressible phases, corresponding to an empty and a fully 
filled lattice (ferromagnetic phases) and a half-filled alternating 
density crystal (anti-ferromagnetic phase), separated from each other 
by compressible phases. In a finite lattice the compressible phases 
show characteristic oscillatory modulations on top of the anti-ferromagnetic
density profile and in density-density correlations. We derive a kink model
which provides simple quantitative explanation of these features. 
To describe the long-range correlations of the system we employ the 
Luttinger liquid theory with the relevant Luttinger parameter $K$ obtained 
exactly using the Bethe Ansatz solution. We calculate the density-density 
as well as first-order correlations and find excellent agreement with 
numerical results obtained with density matrix renormalization group (DMRG)
methods. We also present a perturbative treatment of the system in higher 
dimensions.
\end{abstract}

\pacs{37.10.Jk, 
  03.75.Lm, 
  05.30.Jp, 
  75.10.Jm  
}

\maketitle

\section{Introduction}

Various idealized models describing many-body quantum systems on a lattice, 
such as the Heisenberg spin and Hubbard models, have been 
widely studied for decades in condensed matter physics \cite{SolStPh,MRRrev}. 
With the recent  progress in cooling and trapping bosonic and fermionic 
atoms in optical lattices \cite{OptLatRev}, some of these models can 
now be realized in laboratory with unprecedented accuracy---the 
Hubbard model being a case in point \cite{OL-Hub}. 
Implementing more general models, e.g., extended Hubbard or asymmetric 
spin models, with atoms in optical lattice potentials is, however, more
challenging but potentially very rewarding. The purpose of the present 
paper is to study an experimentally relevant situation realizing
the extended Hubbard model or, equivalently, an anti-ferromagnetic 
$XXZ$ model in the Ising-like phase with cold neutral atoms in a 
deep optical lattice potential.

We consider an optical lattice realization of the Bose-Hubbard 
model with strong on-site attractive interaction between the atoms. 
Specifically, we study a situation when each site of the lattice 
is loaded with either zero or two atoms. Experimentally, this can
be accomplished by adiabatically dissociating a pure sample of 
Feshbach molecules in a lattice with at most one molecule per 
site \cite{feshmols,KWEtALPZ}. The on-site attractive interaction 
then results in the formation of attractively-bound atom pairs 
\cite{molmer,MVDP}---``dimers'',--- whose repulsive analog was 
realized in a recent experiment \cite{KWEtALPZ}. 

For strong atom-atom interaction, either attraction or repulsion, 
the dimer constituents are well co-localized \cite{MVDP}, and an 
ensemble of such dimers in a lattice can be accurately described by
an effective Hamiltonian which has the form of a spin-$\frac{1}{2}$
$XXZ$ model with Ising-like anisotropy. The derivation of the 
effective Hamiltonian is given in \cite{PSAF}, 
where we have also discussed its properties for the case of 
repulsive atom-atom interactions. Since the resulting
nearest-neighbor attraction of dimers dominates the
kinetic energy, it causes the formation of minimal surface ``droplets''
of dimers on a lattice below a critical temperature. In the case of 
attractive atom-atom interaction considered here, the interaction 
between the nearest neighbor dimers is a strong repulsion. We then 
find that the ground state of the system of dimers in a grand canonical
ensemble exhibits incompressible phases, corresponding to an empty and 
a fully filled lattice as well as a half-filled alternating density crystal.
These phases are separated from each other by compressible phases.

We calculate numerically and analytically the ground state phase diagram
for this system in one dimension (1D). The critical points can be obtained 
with the help of the Bethe Ansatz making use of the correspondence to 
the $XXZ$ model \cite{Yang-PR-1966}. In a finite lattice and close to 
half filling, the compressible phases show characteristic oscillatory 
modulations on top of the anti-ferromagnetic density profile. 
A simple kink model is derived which explains the density profiles 
as well as number-number correlations in the compressible phases. 
The long-range correlations of the dimer system show a Luttinger liquid
behavior. We calculate the amplitude and density correlations in a finite 
system from a field theoretical model, which show excellent agreement with
the numerical data. The corresponding Luttinger parameter is obtained 
by solving the Bethe integral equations. We then proceed to a 
perturbative treatment of the system in higher dimensions. Finally,
we briefly discuss the implications of tunable nearest-neighbour 
interactions.

\section{Effective dimer model}

We consider attractively-bound dimers on a $d$-dimensional isotropic 
lattice. Because of the strong on-site atom-atom interaction $U<0$, 
it is energetically impossible to break the dimers, which effectively
play the role of hard core bosons on the lattice. Via a second order 
process in the original atom hopping $J$, the dimers can tunnel to 
neighboring sites with the rate $\tJ \equiv -2 J^2/U > 0$ and carry 
nearest neighbor interaction fixed at $4 \tJ$. The effective 
Hamiltonian for the system has been derived in \cite{PSAF},
\be
\hat H_{\mathrm{eff}} = \sum_j\big(2 \ve_j + U - 2 d \tJ \big) \hat{m}_j
- \tJ \sum_{\expv{j,i}} \hat{c}^\dagger_{j} \hat{c}_{i} 
+ 4 \tJ \sum_{\expv{j,i}} \hat{m}_j \hat{m}_i , \label{eq:dimHam}
\ee
where $\hat{c}^\dagger_{j}$ and $\hat{c}_{j}$ are the creation and annihilation
operators and $\hat{m}_j =  \hat{c}_{j} \hat c_{i}$ is the number operator for 
a dimer at site $j$.  
In the first term of Eq.~(\ref{eq:dimHam}), the local 
potential energy $2 \ve_j $ of the pair of atoms is modified by an 
additional ``internal energy'' of the dimer $\big(U - 2 d \tJ \big)$,
which is negative for attractive interactions, so that the effective
local chemical potential is given by $\mu_j = |U| + 2 d \tJ -2 \ve_j $.
The kinetic energy of 
one dimer described by the second term of Eq.~(\ref{eq:dimHam}) spans 
the interval $[-2 d \tJ, 2 d \tJ]$ corresponding to a Bloch band
of a $d$ dimensional square lattice. In comparison, bringing a pair
of dimers to neighboring sites requires an energy of $8\tJ$ due to 
the strong repulsive interaction in the last term. 

Since the dimers are effectively hard-core bosons, it is possible to 
map the above Hamiltonian onto a spin system. Mapping between bosons 
and spin operators is given by the well known Holstein Primakoff 
transformation  
\bea
\hat S^z_j & = & \hat m_j - 1/2, \nn \\
\hat S^+_j & = & \hat{c}^\dagger_j \, \sqrt{ 1- \hat m_j} , \\
\hat S^-_j & = & \sqrt{1-\hat m_j} \, \hat{c}_j , \nn
\eea
which preserves the SU(2) commutation relations exactly.  
Since double occupancy is forbidden, $m_j=0$ or $1$, the factor 
$\sqrt{1-\hat m_j}$ is zero for an occupied site and unity for 
an empty site. Therefore, we simply have $\hat S^+_j= \hat{c}_j^\dagger$ 
and $\hat S^-_j  =  \hat{c}_j$, so that the equivalent 
Hamiltonian  is given by 
\be
\hat H_{\mathrm{spin}}/\tJ= - \sum_{\expv{j,i}} 
\l( \hat S^+_i \hat S^-_{j} - 4 \hat S^z_i \hat S^z_{j} \r) 
+  \sum_{j} h_j \hat S_j^z , \label{spinham}
\ee
with an effective field of $h_j = (2 \ve_j+U)/\tJ+ 6 d$. 
This is the anti-ferromagnetic $XXZ$ spin model with a fixed 
anisotropy of 4, i.e., the model is in the gapped Ising-like phase. 
A given total number $N$ of dimers in a lattice of $N_s$ sites corresponds,
in the spin model, to a fixed total magnetization $M_z = N - \hlf N_s$.

\section{Dimer system in one dimension}

It is now clear that the behaviour of the dimer system in one dimension 
can be determined via isomorphic mapping of Eq.~(\ref{eq:dimHam}) 
onto the 1D integrable $XXZ$ model in a uniform field,  
\be
\hat H_{XXZ}= - \sum_{j=1}^{N_s-1} 
\l( \hat S^x_j \hat S^x_{j+1} +\hat S^y_j \hat S^y_{j+1} 
- \Delta \hat S^z_j \hat S^z_{j+1} \r) 
+ h \sum_{j=1}^{N_s} \hat S_j^z  , \label{eq:xxzHam}
\ee
where $\Delta$ is the anisotropy of the spin-spin interaction.

\subsection{Ground-state phase diagram}
\label{phasediag}

An important general feature of the dimer model in Eq.~(\ref{eq:dimHam}) 
is that the ratio of interaction to kinetic energy has a fixed value larger
than one. As a consequence, the ground-state of the system is interaction 
dominated giving rise to interesting correlation properties.

In a homogeneous system, 
the ground-state of the system depends only the overall
chemical potential $\mu/\tJ$. 
The corresponding phase diagram can be completely mapped out in an 
experiment by adding a shallow external trapping potential with 
sufficiently small confinement such that the local density approximation
is valid and $\mu = |U| + 2 \tJ -2 \ve_j $  does not change significantly 
over many lattice sites.
Then different regions in the trap would correspond to different 
chemical potentials. 

\begin{figure}[t]
\centerline{\includegraphics[width=0.48\textwidth]{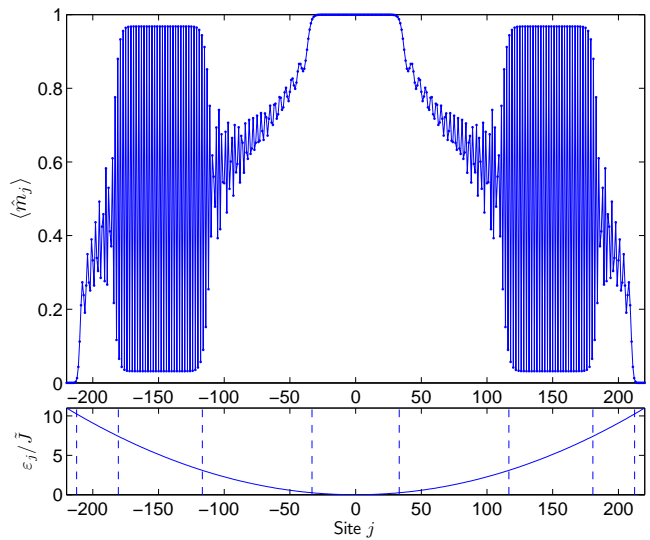}}
\caption{(Color online) Density of dimers in a 1D lattice and 
additional harmonic confinement potential obtained from DMRG 
simulation, with $\mu_j=18.5\tJ -2\epsilon_j$ and $\epsilon_j/\tJ=j^2/4400$. 
One clearly identifies the incompressible phase with homogeneous filling
of $\expv{\hat{m}_j}=1$ in the trap center, and two AF phases, 
separated by compressible intermediate regions.}
\label{fig:AF_trap}
\end{figure}

In Fig.~\ref{fig:AF_trap} we plot the density of dimers 
in a one-dimensional lattice and an additional harmonic trapping 
potential obtained by numerical DMRG calculations \cite{DMRG}.
One clearly recognizes three types of regions: In the trap center, where 
the local chemical potential is largest, there is a unit filling of dimers.
Separated by a spatial region of monotonously decreasing average filling 
follows a region where the filling is exactly one half and the dimers form
a periodic pattern with period 2 and almost maximum modulation depth.
In this region, the dominant effect is the nearest neighbour repulsion 
$4 \tJ > 0$ of Eq.~(\ref{eq:dimHam}). Towards the edge of the dimer 
cloud, the average density decreases again monotonously to zero. 
In terms of the equivalent spin system, the central region corresponds to 
a gapped phase of full spin polarization caused by a large negative 
effective magnetic field. The region of exactly one half average filling 
corresponds to another gapped phase with anti-ferromagnetic (AF) order 
induced by the strong, Ising-like interaction $- 4 \hat S^z_j \hat S^z_{j+1}$
of Eq.~(\ref{spinham}). The intermediate regions are compressible.

The critical values of the chemical potential for the transitions 
between compressible and incompressible phases in 1D are known from 
the work of Yang and Yang \cite{Yang-PR-1966} on the $XXZ$ model 
of Eq.~(\ref{eq:xxzHam}). For the parameters of the present system,
we have
\besa
\mu_\ds / \tJ &=& -2 ,  \label{eq:mu-ds} \\
\mu_{\mathrm{AF-}} /\tJ &=& 8 - 2\sqrt{15} \sum_{n=-\infty}^{\infty} 
\frac{(-1)^n}{\cosh \big( n\, {\arccosh (4)} \big)} \nn \\
&\approx & 3.68361 \ldots , \label{eq:mu-exact-AF-} \\
\mu_{\mathrm{AF+}}/\tJ &=& 8 + 2\sqrt{15} \sum_{n=-\infty}^{\infty} 
\frac{(-1)^n}{\cosh \big( n\, {\arccosh (4)} \big)} \nn \\
&\approx &  12.31638 \ldots ,  \label{eq:mu-exact-AF+} \\
\mu_\us/\tJ &=& 18 . \label{eq:mu-us} 
\eesa
These values agree very well with those obtained from exact 
diagonalization on a small homogeneous lattice with $N_s=10$ sites
and periodic boundary conditions, as well as DMRG simulation with 
up to $N_s=300$ and open boundary conditions. They also match 
the different regions of Fig.~\ref{fig:AF_trap}.

\subsection{Mott-insulating phases}
\label{sec:mott}

In the language of spin Hamiltonian, phases with zero ($N=0$) or full 
($N=N_s$) filling correspond to ferromagnetic phases with a simple form 
of the ground state 
\bea
\ket{\psi_\ds} &=& \ket{\ds,\ds,\ds,\ldots,\ds} , \label{eq:dsS} \\
\ket{\psi_\us} &=& \ket{\us,\us,\us,\ldots,\us} . \label{eq:usS}
\eea
Particle-hole excitations are not possible in the Mott-insulating state
(\ref{eq:usS}), while inserting a particle into (\ref{eq:dsS}) or 
removing one from (\ref{eq:usS}), corresponding to flipping a spin, 
carries finite energy cost given by Eqs.~(\ref{eq:mu-ds}) and (\ref{eq:mu-us}).
Hence these phases are incompressible.

For half filling ($N= \hlf N_s$) the situation corresponds most closely 
to an AF phase. However, in this case the simple N\'eel state  
\be
\ket{\psi^{(0)}_{\mathrm{AF}}} = \ket{\ldots,\ds,\us,\ds,\us,
\ds,\us,\ds,\us, \ldots } , \label{neel}
\ee
is not an exact eigenstate of the full Hamiltonian $\hat H_{\mathrm{eff}}$ 
in (\ref{eq:dimHam}). Rather, $\ket{\psi^{(0)}_{\mathrm{AF}}}$ is an eigenstate
of Hamiltonian $\hat H_{\mathrm{eff}}^{(0)} \equiv \hat H_{\mathrm{eff}} 
- \hat{H}_{\mathrm{hop}}$ without the hopping term 
$\hat{H}_{\mathrm{hop}}=- \tJ \sum_{j} (\hat{c}^\dagger_{j+1} \hat{c}_j
+ \hat{c}^\dagger_{j} \hat{c}_{j+1})$. Due to $\hat{H}_{\mathrm{hop}}$
a dimer can tunnel from an occupied site to a neighboring empty site,
which in terms of the N\'eel state (\ref{neel}), corresponds to 
flipping two neighboring spins, resulting in a state of the form
\be
\ket{\psi_j^{(1)}} = \ket{\ldots,\ds,\us,\ds,\ds_j,\us_{j+1},
\us,\ds,\us,\ldots} . \label{excited}
\ee 
If we assume periodic boundary conditions and an even number of lattice
sites $N_s$, there are $j=1,\ldots,N_s$ different states (\ref{excited}),
one for each link where two neighboring spins can be flipped. 
Each of those states $\ket{\psi^{(1)}_j}$ has a larger repulsive
(Ising) interaction energy $E^{(1)}_j$, which is increased by $8\tJ$ 
relative to energy $E^{(0)}_{\mathrm{AF}}$ of state 
$\ket{\psi^{(0)}_{\mathrm{AF}}}$. 
It is tempting to treat the the smaller hopping 
$\hat{H}_{\mathrm{hop}}$ as perturbation with respect to 
$\hat H_{\mathrm{eff}}^{(0)}$, but unfortunately already the first order correction
carries a contribution from all $N_s$ possible states in Eq.~(\ref{excited}).  In higher
order perturbation theory the number of contributing states 
increases with higher powers in $N_s$, so that the perturbation series diverges in 
the thermodynamic limit.

However, if we are interested in local observables, such as the density 
in Fig.~\ref{fig:AF_trap}, it is possible to restrict the perturbation only to those 
hopping terms which change the value of the density at a particular point.
In particular, in order to calculate the ground state expectation value of 
$\langle \psi_{\mathrm{AF}}| S_j^z|\psi_{\mathrm{AF}}\rangle$, we have to make the 
following ansatz for the ground state $|\psi_{\mathrm{AF}}\rangle$
\bea
\ket{\psi_{\mathrm{AF}}} & \approx & 
\ket{\psi^{(0)}_{\mathrm{AF}}} + \sum_{i=j-1}^{j} 
\frac{\ket{\psi_i^{(1)}} \bra{\psi_i^{(1)}} \hat{H}_{\mathrm{hop}}
\ket{\psi^{(0)}_{\mathrm{AF}} }}
{E^{(0)}_{\mathrm{AF}} - E^{(1)}_i} \nn \\ 
& = & \ket{\psi^{(0)}_{\mathrm{AF}}} 
+ \frac{1}{8} \left(\ket{\psi_{j-1}^{(1)}}+\ket{\psi_{j}^{(1)}}\right) , \label{af-gs}
\end{eqnarray}
which can be normalized by a factor of $1/\sqrt{1+1/32}$. 
This state is in general a bad approximation to the ground state, but 
it describes very well which terms in the Hamiltonian affect the 
local density at site $j$, since higher order hopping only contributes 
$1/64$ or less.  Accordingly, the local density is given by 
\bea
\langle \psi_{\mathrm{AF}}| S_j^z|\psi_{\mathrm{AF}}\rangle \approx
(-1)^j \frac{32}{66}\left(1-\frac{1}{32}\right) = (-1)^j \frac{31}{66}
\eea
which corresponds to a deviation of about $0.03$ from perfect alternating
order.
Our numerical results for homogeneous systems show a deviation of about $0.032$ which is 
in very good agreement with the prediction.
Even though the half filling state always implicitly contains 
excitations of type (\ref{excited}), the removal or addition of 
a particle still costs relatively large energy given by 
Eq.~(\ref{eq:mu-exact-AF-}) or (\ref{eq:mu-exact-AF+}), 
which makes the AF phase incompressible.

\subsection{Properties of compressible phases}
\label{sec:propcp}

In the remainder of this Section, we examine the compressible phases, 
mainly in the vicinity of the antiferromagnetic phase, using 
two different approaches. 
The first is perturbative in nature and relies on the fact that the 
nearest neighbor interaction energy between the dimers exceed the dimer 
hopping energy by a large factor of 8. We show that the system can 
approximately be treated as a non-interacting gas of kinks that behave
like hard-core bosons. 
The second approach aims to describe long-range correlations employing 
the Luttinger-liquid theory. The relevant Luttinger parameter can be 
obtained by Bethe Ansatz solution of the equivalent $XXZ$ spin model.

\subsubsection{Non-interacting kink approximation}
\label{sec:kink}

\begin{figure}[t]
\centerline{\includegraphics[width=0.48\textwidth]{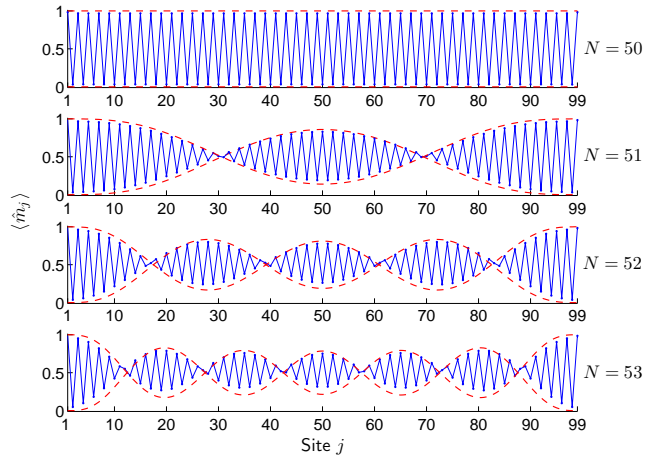}}
\caption{(Color online) Particle density profile in a homogeneous
lattice with $N_s = 99$ sites and hard-wall boundaries, for different 
particle number $N$. For half filling, $N = 50$, the ground state 
has nearly perfect AF order. Adding one, two and three particles
leads to the density-wave modulations with the number of nodes equal 
twice the number of additional particles.}
\label{fig:density}
\end{figure}

In Fig.~\ref{fig:density} we plot the density distribution of dimers
in a homogeneous lattice of $N_s = 99$ sites obtained from DMRG 
simulations with different number of dimers $N$. An infinite (hard-wall) 
confining potential $\epsilon_0=\epsilon_{100}\to +\infty$ has been used, 
which imposes on Hamiltonian~(\ref{eq:dimHam}) open boundary conditions 
with $m_0=m_{100}=0$. Due to the asymmetric coupling at the boundaries, 
the end sites $j=1$ and $j=99$ prefer to be occupied with a particle.  
To accommodate an oscillating density wave, we therefore use odd number 
of lattice sites $N_s$. Note that the open boundary condition for the 
particles in Eq.~(\ref{eq:dimHam}) corresponds to an additional effective 
edge field $h_1 = h_{99} = -2$ for the spins in the $XXZ$-model 
of Eq.~(\ref{spinham}), which has the analogous effect of polarizing 
both end spins up.

As seen in Fig.~\ref{fig:density}, the ground state for $N=50$ exhibits
density oscillations corresponding to the AF N\'eel order, 
up to the small correction discussed in Sec.~\ref{sec:mott}. Adding 
particles leads to modulated density distribution, with the envelope 
of modulation having regularly spaced nodes whose number is equal to 
twice the number of additional particles. In the following we will 
provide a simple theoretical understanding for this effect. 

Without the small hopping term $\hat{H}_{\mathrm{hop}}$, the ground 
state of Hamiltonian~(\ref{eq:dimHam}) for half filling is the AF 
state $\ket{\psi^{(0)}_{\mathrm{AF}}}$ of Eq.~(\ref{neel}), which is twofold 
degenerate. The AF order with period 2 effectively doubles
the size of the unit cell. Adding then a particle to 
$\ket{\psi^{(0)}_{\mathrm{AF}}}$ costs exactly an energy of 
$(h + 8)\tJ$, resulting in state
\[
\ket{\ldots,\us,\ds,\us_j,\us_{j+1},\us,\ds,\us,\ds,\us, \ldots} , 
\]
which is energy-degenerate with any state of the form
\be
\ket{ \psi_{\mathrm{AF}+1}^{(0)}} 
= \ket{\ldots,\us,\ds,\us_j,\us,\ds,\ldots,\ds,\us_{j'},\us,\ds, \ldots} . 
\label{eq:AF+1}
\ee
Hence, the additional particle causes effective domain walls, 
which can be placed anywhere in the system and play the role 
of mobile kinks at positions $j$ and $j'$ between 
AF regions with different orientation. 
Note that without hopping any number of particles above half 
filling can be created at the critical field $h_c^{(-)} = - 8$
and placed in an arbitrary arrangement as long as no two 
neighboring lattice sites are empty. In other words, at $h_c^{(-)}$
we have a huge degeneracy of states with any magnetization $M_z \geq 0$
corresponding to arbitrary arrangement of antiferromagnetic regions and 
spin-up ferromagnetic regions. The analogous statement is also true 
at the upper critical field $h_c^{(+)} = 8$, where the degenerate 
subspace is defined as states with $M_z \leq 0$ where no two neighboring 
spins may point up. This degeneracy implies that without hopping the 
transition from the antiferromagnetic incompressible phase to the 
ferromagnetic incompressible phases is infinitely sharp at the effective 
critical magnetic fields $h_c^{(\pm)}$. As we will see below, however, 
the hopping lifts this degeneracy and therefore is crucial for the 
stability of compressible phases over finite ranges of field $h$ as 
observed in Fig.~\ref{fig:AF_trap}.

\begin{figure}[t]
\centerline{\includegraphics[width=0.35\textwidth]{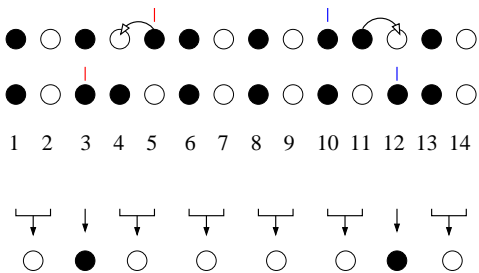}}
\caption{(Color online) Top: 1D chain with one particle added to the 
AF state creating a pair of odd (red) and even (blue) kinks.
The hopping Hamiltonian $\hat{H}_{\mathrm{hop}}$ leads to a motion of 
the odd and even kinks on odd or even sites, respectively. 
Interchange of odd- and even-site kinks is not possible. 
Bottom: mapping onto an effective lattice with lattice constant 2.}
\label{fig:defect-hopping}
\end{figure}

The hopping $\hat{H}_{\mathrm{hop}}$ is also responsible for 
the modulated wave patterns seen in Fig.~\ref{fig:density}. 
Starting from the AF state in Eq.~(\ref{neel}), we now insert more 
and more particles each producing a pair of kinks. The states 
in Eq.~(\ref{eq:AF+1}) can be considered as AF states with a pair 
of kinks, one at even sites and one at odd sites; e.g., state 
$\ket{ \ds_1,\us_2,\ds_3,\us_4,\us_5,\us_6,\ds_7,\us_8,\ds_9,\ldots}$
has kinks at sites 4 and 5, while state
$\ket{ \ds_1,\us_2,\ds_3,\us_4,\us_5,\ds_6,\us_7,\us_8,\ds_9,\ldots}$
has kinks at sites 4 and 7. $\hat H_{\rm hop}$ has non-vanishing matrix 
elements within the subspace of energy degenerate states with fixed 
number of additional particles. Within this manifold of states, hopping 
of the additional particle corresponds to free motion of kinks, 
wherein an even-site kink moves only on even sites and an odd-site 
kink on odd sites, as illustrated in the top part of 
Fig.~\ref{fig:defect-hopping}. Furthermore, the even and odd site chain 
kinks cannot exchange their relative order. Note that hoping of a 
particle surrounded by two empty sites is energetically suppressed.

Given a fixed number of additional particles or holes, the motion 
of the corresponding kinks is equivalent to the motion of hard-core 
bosons in an effective lattice with lattice constant 2. 
To see this, consider the case of $q$ additional particles on top of the
half filled lattice; the opposite case of holes follows from the 
particle--hole symmetry. Let the positions of the kinks be 
$j_1<j_2<\ldots <j_{2q}$. If $j_1$ is even (odd) then 
$j_3,j_5,j_7,\ldots$ are also even (odd) and 
$j_2,j_4,j_6,\ldots$ are odd (even). We now perform a mapping
onto a new lattice which we call the kink lattice. 
The quasi-position $k_n$ of the $n$th kink is then
\begin{align}
k_n = \begin{cases} 
\frac{j_n + n-1}{2} & \mbox{ if $j_1$ is even} \\
\frac{j_n + n  }{2} & \mbox{ if $j_1$ is odd}
\end{cases} .
\end{align}
This mapping is illustrated in the lower part of Fig.~\ref{fig:defect-hopping}. 

Evaluating the matrix elements of the hopping Hamiltonian $\hat H_{\rm hop}$
in the subspace of states with constant number of kinks, we find that 
the latter can be treated as hard-core bosons or non-interacting 
fermions on the kink lattice, only if we consider the absolute value of 
the wavefunction. The corresponding hopping strength on the period-2
lattice is again $\tJ$. The exchange symmetry cannot be determined
straightforwardly and therefore we employ this approximation only to 
determine the density distribution of dimers. For simplicity we choose 
the fermionic exchange symmetry.

Let us assume that the lattice is large and consider a particle filling 
close to the antiferromagnetic case. In this limit, the kinks can be 
regarded as moving on a continuum. This means that the dynamics of the 
kinks can now be determined by solving the Schr\"odinger equation for 
non-interacting fermions. For $N=\hlf N_s +1$, i.e., one additional 
particle, we have a pair of kinks whose ground-state wavefunction is
\be
\Psi_2(x_1,x_2) = \frac{\sqrt{2}}{L} 
\left[\sin \frac{\pi x_1}{L}  \, \sin \frac{2\pi x_2}{L}  
- \sin \frac{\pi x_2}{L} \,  \sin \frac{2 \pi x_1}{L} \right] ,
\ee
where $L = \hlf N_s + 1$ is the length of the kink lattice.
The left-most kink shall move on the odd sites. A particle is sitting 
on an even site $j$ if and only if one chain kink is to the left of $j$.
Thus the density of particles on the even sites is
\be
\expv{\hat{m}(x)} = 2 \int_0^{x} \!\! d y_1 \! 
\int_x^{L}\!\! d y_2 \, \Psi_2^*(y_1,y_2) \Psi_2(y_1,y_2) .
\label{eq:2qdens}
\ee
The prefactor of two emerges here because the integral occurs twice 
with interchanging the roles of $y_1$ and $y_2$. Although straightforward,
we do not give the analytic expression of Eq.~(\ref{eq:2qdens}) since 
it is rather long. At the odd sites we get accordingly
\bea
1- \expv{\hat{m}(x)} &=& 
\int_0^{x}\!\! d y_1 \! \int_0^{x} \!\! d y_2 \, \Psi_2^*(y_1,y_2)\Psi_2(y_1,y_2)
\nn \\ & & +
\int_x^{L}\!\! d y_1 \! \int_x^{L} \!\! d y_2\, \Psi_2^*(y_1,y_2) \Psi_2(y_1,y_2)
. \qquad
\eea
With $q$ additional particles, the fermionic ground state wavefunction 
for $2q$ kinks is
\be
\Psi_{2q} (x_1,\ldots,x_{2q}) = 
\sum_P \frac{\sgn (P)}{\sqrt{(2q)!}} \prod_{n=1}^{2q} \phi_{P(n)}(x_n) ,
\ee
where the sum is over all permutations $P$ of numbers $\{1,2,3,\ldots,2q\}$ and
\[
\phi_n(x)=\sqrt{\frac{2}{L}} \, \sin \frac{\pi n x}{L} ,
\]
with $L = \hlf N_s + q$. This results in the density distribution
\begin{multline}
\expv{\hat{m}(x)} = \sum_{k=0}^{q-1}\sum_{P,Q}
\left[ \frac{\sgn (P) \, \sgn (Q)}
{(2k+1)!(2q-2k-1)!} \right. \\  
\left. \times \prod_{n=1}^{2k+1} I \big( 0,x,P(n),Q(n) \big) \!\!
\prod_{n=2k+2}^{2q} \!\! I \big( x,L,P(n),Q(n) \big ) \right] , 
\label{eq:density}
\end{multline}
where $Q$ denotes the permutations of $\{1,2,3,\ldots,2q\}$, and
\[
I(a,b,n,m) = \int_a^b \! d x \, \phi^*_n(x) \phi_m(x) , 
\]
with $n,m\in\{1,2,3,\ldots,2q\}$. In Eq.~(\ref{eq:density}) we have taken 
into account that there are $\frac{(2q)!}{(2q-2k-1)!(2k+1)!}$ possibilities 
of choosing $2k+1$ kinks to the left of $j$. 

The dashed red lines in Fig.~\ref{fig:density} show the analytic results 
for the particle density in a box potential with the lattice filling 
slightly above one half obtained from the kink approximation. The agreement
with the numerical DMRG data is rather good. The kink model also explains
in a very intuitive way the pairwise appearance of nodes with adding 
every particle to the lattice. 

\begin{figure}[t]
\centerline{\includegraphics[width=0.48\textwidth]{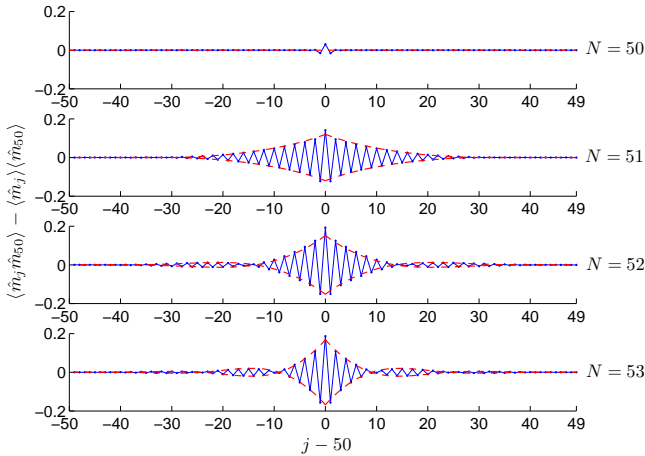}}
\caption{(Color online) Particle density-density correlations in a 
homogeneous lattice with $N_s = 99$ sites and hard-wall boundaries,
for different particle number $N$. The blue lines correspond to 
numerical DMRG results, the red dashed lines to the predictions 
of the kink approximation.}
\label{fig:density-density}
\end{figure}

Particle number correlations can be derived in the same manner.
For two even sites at positions $j_1$ and $j_2$, the configurations 
contributing to the correlations correspond to an odd number of 
particles to the left of $j_1$, an even number of particles between
$j_1$ and $j_2$, and an  even number of particles to the right of $j_2$. 
The particle density-density correlations are then given by
\begin{multline}
\expv{ \hat m(x) \hat m(y)} = \!\!
\sum_{k_1,k_2,k_3=0}^{\genfrac{}{}{0pt}{}{k_1+k_2+k_3}{\leq (q-1)}}
\sum_{P,Q} \left[ \frac{\sgn (P) \, \sgn (Q)}
{(2k_1+1)!(2k_2)!(2k_3+1)!} \right. \\ 
\times \prod_{n=1}^{2k_1+1} I\big( 0,x,P(n),Q(n) \big) \!\!
\prod_{n=2k_1+2}^{2k_1+2k_2+1} \!\! I\big( x,y,P(n),Q(n) \big) \\ 
\left.  \times  \!\!\!\!\!\! \prod_{n=2k_1+2k_2+2}^{2q} \!\! 
I \big( y,L,P(n),Q(n) \big) \right]
, \quad \mathrm{for} \quad x < y . \label{eq:correlation}
\end{multline}

In Fig.~\ref{fig:density-density} we plot the density-density correlations 
obtained from DMRG calculations (blue solid line) and the kink model 
(dashed red lines), displaying very good agreement. We finally note that 
within the approximation of non interacting kinks, first order correlations
exist only between neighboring sites. This perturbative model therefore 
can not accurately describe such correlations.

\subsubsection{Field theoretical approach}
\label{sec:ftapr}

The spin chain equivalent to the dimer Hamiltonian~(\ref{eq:dimHam}) 
in 1D is given by Eq.~(\ref{eq:xxzHam}) with $\Delta=4$ and open boundary 
conditions. At zero magnetization $M_z=0$, the $XXZ$ model~(\ref{eq:xxzHam})
is gapped, since $\Delta > 1$. However, as described in Sec.~\ref{phasediag},
the gap can be closed by a field between the two critical values, 
$h_c^{(-)} < h < h_c^{(+)}$. In other words, the system is critical for 
any nonzero magnetization away from the fully magnetized case. 
In this regime, the leading low-energy effective theory is 
a Luttinger liquid with two parameters, the spin velocity $v$ 
and Luttinger parameter $K$. These are functions of the magnetization 
per site $s^z= M_z/N_s$ and anisotropy $\Delta$ \cite{hal80}
(which for the particular dimer model here is fixed, $\Delta=4$).
 
\begin{figure}[b]
\centerline{\includegraphics[width=0.4\textwidth]{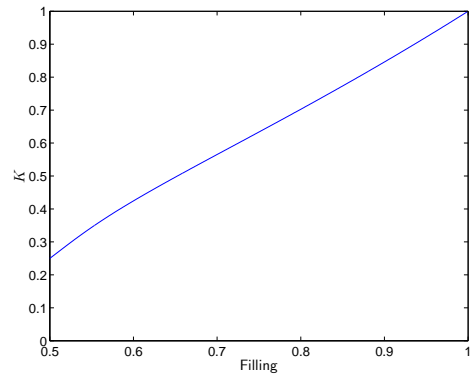}}
\caption{Dependence of the Luttinger parameter $K$ on 
the mean lattice filling $N/N_s$, for $\Delta = 4$.}
\label{fig:Luttinger-K}
\end{figure}

In order to calculate correlation functions, we first derive
the Luttinger parameter $K (s^z)$ from the exact solution \cite{bog86}.
We write
\bea
K=\xi^2(b), \label{KLut}
\eea
where the function $\xi(x)$ is determined by the integral equation
\bea
\xi(x) = 1 + \int_{-b}^b \kappa(x-y) \xi(y) \, d y , \label{xifnc}
\eea
with the kernel 
\[
\kappa(x) = \frac{1}{\pi} 
\frac{\sinh 2\eta}{\cos 2 x - \cosh 2\eta}, \qquad \Delta = \cosh\eta > 1.
\]
The parameter $b$ in Eqs.~(\ref{KLut}) and (\ref{xifnc}) is implicitly 
defined through
\besal{szrho}
s^z &=& \frac12 - \int_{-b}^b \rho(x) \, d x ,  \\
\rho(x) &=& d(x)+\int_{-b}^b \kappa(x-y) \rho(y) \, d y, 
\eesa
where 
\[
d(x) = \frac{1}{\pi} 
\frac{\sinh \eta}{\cos 2 x - \cosh \eta} , \qquad \Delta = \cosh\eta > 1 .
\]
Equations (\ref{KLut})-(\ref{szrho}) are solved numerically by 
discretizing the integral and inverting the resulting matrix equation. 
Figure \ref{fig:Luttinger-K} shows the function $K (s^z)$ for $\Delta = 4$. 

\begin{figure}[t]
\centerline{\includegraphics[width=0.48\textwidth]{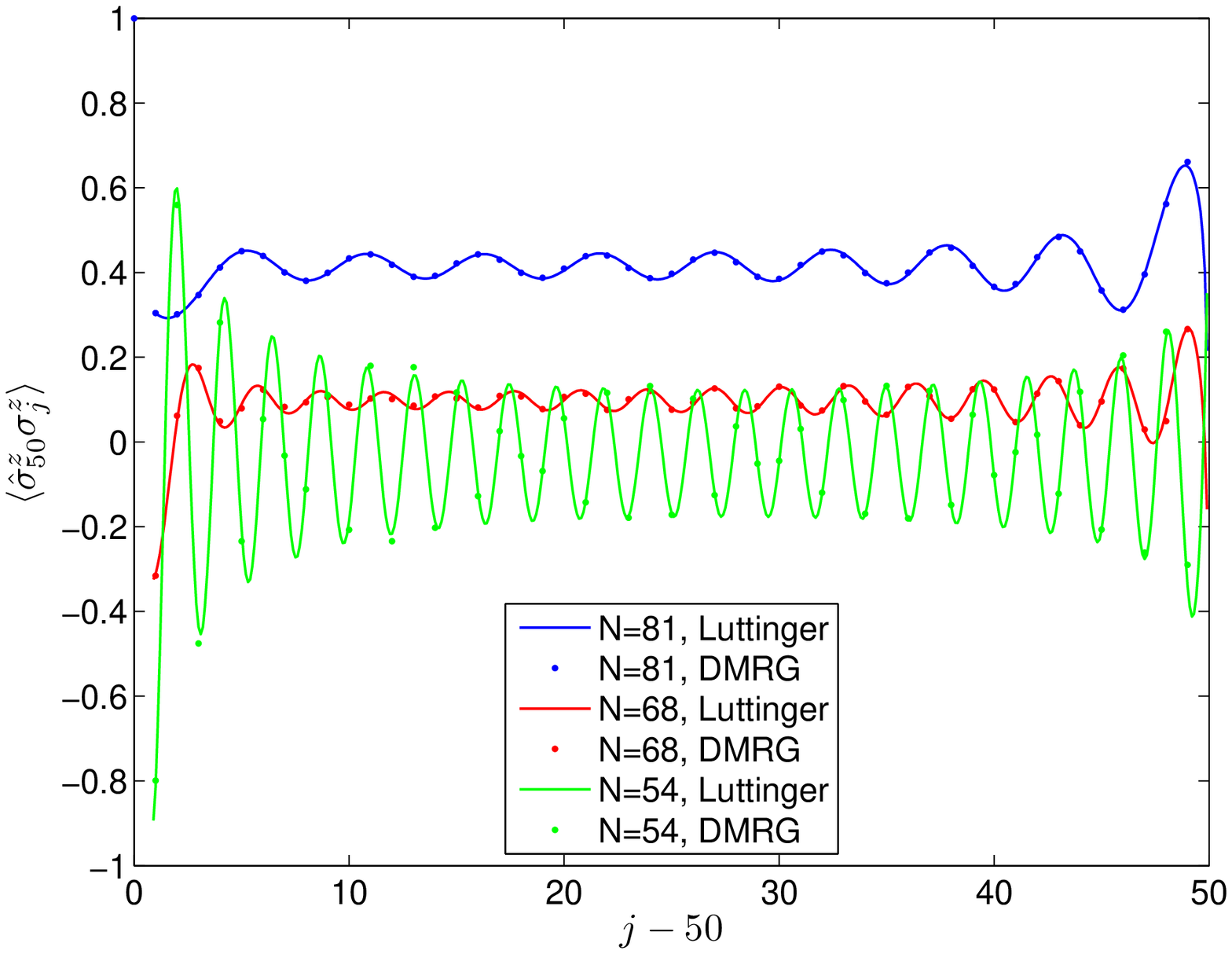}}
\caption{(Color online) $\expv{ \hat S^z(x) \hat S^z(50) }$ correlations 
obtained from the DMRG (dots) and the Luttinger-liquid approximation 
(solid lines).}
\label{fig:correlations}
\end{figure}

\begin{figure}[t]
\centerline{\includegraphics[width=0.48\textwidth]{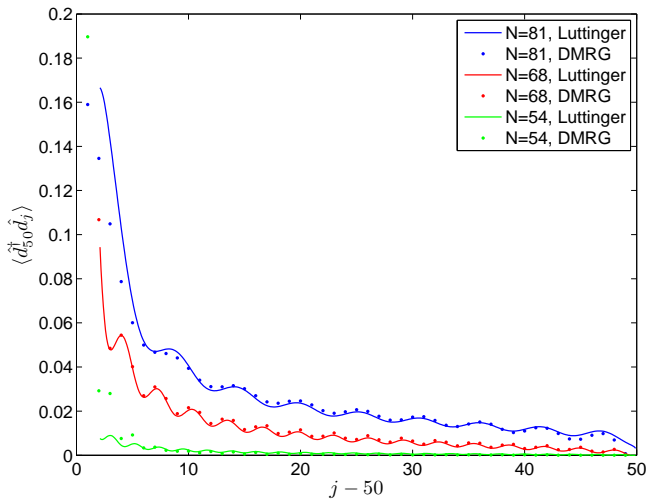}}
\caption{(Color online) $\expv{\hat S^+(x) \hat S^-(50)}$ correlations 
obtained from the DMRG (dots) and the Luttinger-liquid approximation 
(solid lines).}
\label{fig:correlationspm}
\end{figure}

Within the Luttinger liquid approach, one- and two-point correlation 
functions can be calculated using the standard mode expansion of bosonic
fields \cite{egg_book06} for open boundary conditions \cite{egg92,caz04}. 
Then the spin-spin correlation function in the ground state reads
\begin{widetext}
\bea
\expv{\hat S^z(x) \hat S^z(y) } &=& (s^z)^2 
- B \, \frac{K}{8(N_s+1)^2} 
\l( \frac{1}{ \sin^2 \frac{\pi(x-y)}{2(N_s+1)} }
+ \frac{1} { \sin^2 \frac{\pi(x+y)}{2(N_s+1)}  } \r)
+ C_1 \, \frac{\cos\l[(2k_F+\theta/N_s)x + \varphi_1\r]}
{\l(\sin\frac{\pi x}{N_s+1}\r)^K} \nn \\ & & 
+ C_2 \, \frac{\cos\l[(2 k_F+\theta/N_s)y + \varphi_2\r]}
{\l(\sin\frac{\pi y}{N_s+1}\r)^K}
+ D \, \frac{\cos\l[(2 k_F+\theta/N_s)x+\delta\r]}
{\l(\sin\frac{\pi x}{N_s+1} \, \sin \frac{\pi y}{N_s+1}\r)^K}
\l[ \frac{\sin \frac{\pi(x+y)}{2(N_s+1)}}
{\sin \frac{\pi(x-y)}{2(N_s+1)}}\r]^{2 K} , \label{szsz}
\eea
with the Fermi wavevector $k_F \equiv \pi(1-2 s^z)/2$. Here the amplitudes 
$B,C_{1,2},D$,  the shift $\theta$, and the phases $\varphi_{1,2}, \delta$ result from bosonization of operators on the lattice. We consider  
them as parameters in Eq.~(\ref{szsz}) that are fixed numerically 
by fitting to the DMRG data. The exponents, however, are obtained from the 
Luttinger liquid parameter $K$, which is given by the Bethe Ansatz. 
Figure~\ref{fig:correlations} shows the remarkable agreement between 
the two approaches. Note the shift in the wavevectors of the 
oscillations by a constant $\theta$ that depends on the boundary conditions, the interaction and the magnetization. It has also been observed in the 
context of density oscillations in the open Hubbard model \cite{bed98,soe08}. 

The corresponding result for the first-order correlation function 
in the ground state is
\bea
\expv{ \hat S^+(x) \hat S^-(y) } &=& 
\l[ \frac{\sqrt{ \sin \frac{\pi x}{N_s+1} \, \sin \frac{\pi y}{N_s+1}}}
{\sin \frac{\pi (x+y)}{2(N_s+1)} \, \sin \frac{\pi (x-y)}{2(N_s+1)}}\r]^{1/(2 K)}
\l( B \, \frac{\cos \l[(2 k_F+\theta/N_s)(x-y)+\delta\r]}
{\l(\sin\frac{\pi x}{N_s+1} \, \sin \frac{\pi y}{N_s+1}\r)^K}
\l[\frac{\sin \frac{\pi (x+y)}{2(N_s+1)}}
{\sin \frac{\pi (x-y)}{2(N_s+1)}}\r]^{2 K}\r. \nn \\ & & \l.
+ C_1 \, \frac{\cos\l[(2 k_F+\theta/N_s)x + \varphi_1\r]}
{\l( \sin \frac{\pi x}{N_s+1}\r)^K}
+ C_2 \, \frac{\cos\l[(2 k_F+\theta/N_s)y + \varphi_2\r]}
{\l(\sin\frac{\pi y}{N_s+1}\r)^K}\r)\label{spsm}.
\eea
\end{widetext}
Similarly to Eq.~(\ref{szsz}), the quantities $B,C_{1,2},D, \delta, \phi_{1,2}, \theta$ are considered as fitting parameters. The resulting curves are shown in 
Fig.~\ref{fig:correlationspm}.

\section{Phase diagram in higher dimensions}

We now derive the phase boundaries for the dimer system in two and 
three dimensions. To that end, we employ the strong-coupling approach 
\cite{Freericks-PRB-1996}, wherein the hopping term 
$\hat{H}_{\mathrm{hop}} =- \tJ \sum_{\expv{j,i}} \hat{c}^\dagger_j \hat{c}_i$
of Hamiltonian~(\ref{eq:dimHam}) is treated as small perturbation
with respect to $\hat{H}^{(0)}_{\mathrm{eff}} \equiv 
\hat{H}_{\mathrm{eff}} - \hat{H}_{\mathrm{hop}}$.

\paragraph{Zero-hopping limit.}

Without the hopping, the grand canonical operator for the dimer system
reads
\be
\hat{H}^{(0)}_{\mathrm{eff}} 
= 4 \tJ \sum_{\expv{j,i}} \hat{m}_j \hat{m}_i - \mu \sum_j \hat m_j . 
\ee
In this (formal) limit, the model is isomorphic to the Ising model 
in an external magnetic field and it has two critical points 
\besa
\mu_{\ds}^{(0)}/\tJ &=& 0 , \\
\mu_{\us}^{(0)}/\tJ &=& 16d .
\eesa
For very small values of the chemical potential, 
$\mu < 0$, 
all spins are polarized in the $-z$ direction, which in the dimer 
language corresponds to a state with zero dimers at each lattice site. 
For sufficiently large values of chemical potential, 
$\mu > 16 d \tJ$, all spins are aligned 
in the $+z$ direction, i.e., we have unit filling of the dimer lattice.
Finally, for intermediate values of the chemical potential, 
$\mu_{\ds}^{(0)} < \mu <\mu_{\us}^{(0)}$, the ground state is twofold 
degenerate and has antiferromagnetic order corresponding to a 
``checkerboard-crystal'' lattice of dimers.

\paragraph{Boundaries of ferromagnetic phases.}

When the hopping term $\hat{H}_{\mathrm{hop}}$ is brought into 
the picture, the two critical points extend to two critical regions in 
which the system is compressible. In order to determine the chemical 
potentials at which the transitions between the compressible and 
incompressible ferromagnetic phases take place, we calculate the 
particle and hole excitation energies for zero and full filling 
of a finite lattice with even number of sites $N_s$ using  
Hamiltonian~(\ref{eq:dimHam}). 

In the case of a single dimer in an empty lattice, there is no 
contribution from the interaction energy and we find immediately
without resorting to perturbative treatment
\beann
E(N=0) &=& 0, \\
E(N=1) &=& (U - 2 d \tJ ) - 2d \tJ . 
\eeann
In the filled lattice, each nearest neighbour link contributes
$8 \tJ$ of repulsive interaction energy and we obtain
\beann
E(N = N_s)  &=& (U - 2 d \tJ ) N_s + 8 d \tJ N_s , \\
E(N= N_s-1) &=& (U - 2 d \tJ) (N_s-1) \\ & & 
+ 8 d \tJ (N_s - 2) - 2 d \tJ . 
\eeann
The critical chemical potentials, are then determined by the energy difference
$ E(1) - E(0)$, 
at which we add a dimer to the empty lattice, and 
$ E(N_s) - E(N_s-1)$, at which we add a hole to 
(or remove a dimer from) the filled lattice
\besa
\mu_{\ds} /\tJ&=& -2 d ,  \label{eq:mu-down} \\
\mu_{\us}/ \tJ&=& 18 d . \label{eq:mu-up}
\eesa 

It should be noted that the hopping Hamiltonian does not modify 
the corresponding states in the two insulating phases, i.e., within 
the empty and fully-filled lattice phases there are no fluctuations 
of the dimer number, which is exactly zero or one per site, respectively.

\paragraph{Boundaries of antiferromagnetic phase.}

\begin{figure}[b]
\begin{center}
\includegraphics[width=3.2 cm]{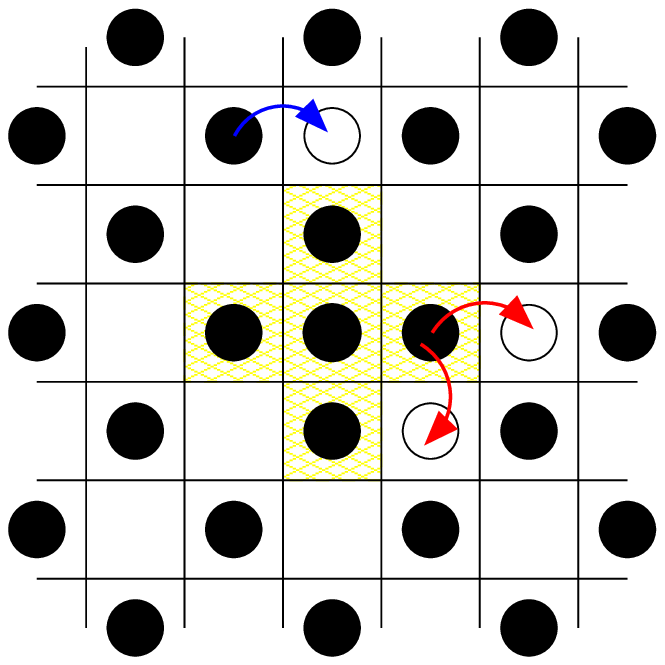}
$\qquad$
\includegraphics[width=3.2 cm]{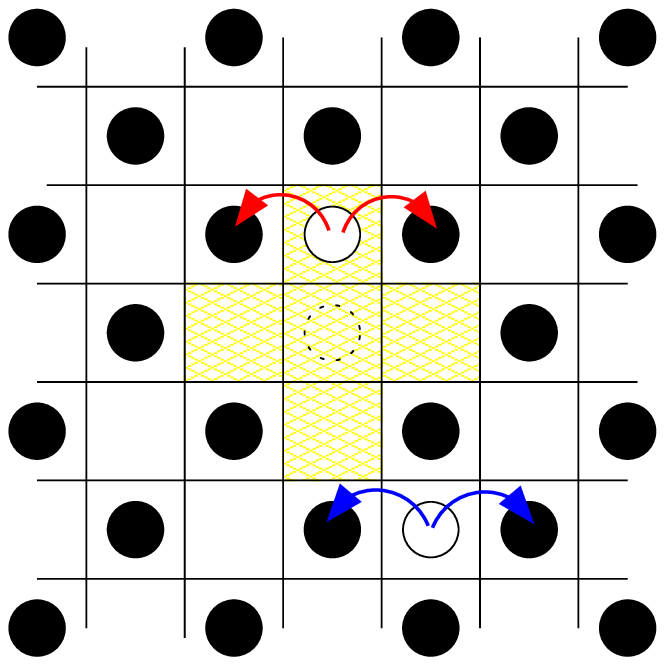}
\end{center}
\caption{(Color online) AF state in 2D lattice with an additional
particle (left) or hole (right). Virtual hopping of particles
(holes) adjacent to the defect (red) and in the bulk (blue) 
lead to different second-order energy contributions. 
Hopping of the additional particle or hole is not allowed.}
\label{fig:AF-defects}
\end{figure}

We now calculate the lower and upper critical chemical potentials 
$\mu_{\mathrm{AF}\mp}$ for the AF phase, up to the second order in 
dimer hopping. At exactly half filling, $N=\hlf N_s$, the ground state
is an almost perfect antiferromagnet with an alternating density 
structure in which there are no nearest-neighbour repulsive junctions
between the dimers. However, due to $\hat{H}_{\mathrm{hop}}$ each dimer
undergoes highly nonresonant transitions to the neighboring empty sites,
whose number is $2d$ (see Fig.~\ref{fig:AF-defects} illustrating the 2D case),
and we obtain
\[
E^{(2)} \big( \hlf N_s \big) = (U - 2 d \tilde J) \hlf N_s - 
\frac{(\tJ)^2}{8 \tJ (2d-1)} \, \hlf N_s \, 2d  , 
\]
where the last term describes the second-order energy shifts 
resulting from the virtual transitions of the dimers.

In the cases of $N=\hlf N_s \pm 1$, the added dimer or dimer hole can not
freely move for $d \geq 2$, since it would require two hopping events, 
as can be seen in Fig.~\ref{fig:AF-defects}. This should be contrasted 
with the 1D situation, wherein adding a dimer or a hole to the AF phase 
creates two mobile kinks, discussed in Sec.~\ref{sec:kink}.
Taking into account the second order corrections due to virtual 
transitions of dimers adjacent to the extra dimer or dimer-hole
(see Fig.~\ref{fig:AF-defects} left or right, respectively),
we obtain
\beann
E^{(2)} \big(\hlf N_s \pm 1 \big) &=& 
(U - 2 d \tJ ) \big( \hlf N_s \pm 1 \big) \\ && 
- \frac{ (\tJ)^2}{8 \tJ (2d-1)} \, \big(\hlf N_s  - 2 d \big) \, 2d \\ &&
- \frac{ (\tJ)^2}{8 \tJ( 2d-2)} \, 2 d \, (2d-1) .
\eeann

The lower $\mu_{\mathrm{AF}-} = E^{(2)} \big( \hlf N_s \big) 
- E^{(2)} \big(\hlf N_s - 1 \big)$ and upper 
$\mu_{\mathrm{AF}+} = E^{(2)} \big( \hlf N_s + 1\big) 
- E^{(2)} \big(\hlf N_s \big)$ critical chemical potentials 
for the AF phase in $d \geq 2$ dimensions are then given by
\besa
\mu_{\mathrm{AF}-} / \tJ &=&  \frac{d}{4 (2d-1)(2d-2)} , \\
\mu_{\mathrm{AF}+} / \tJ &=& 16 d  - \frac{d}{4 (2d-1)(2d-2)} .
\eesa

\section{The role of anisotropy $\Delta$ }

The effective Hamiltonian (\ref{eq:dimHam}) has a fixed relation of 
the nearest neighbour interaction to hopping, which results in a fixed 
Ising like anisotropy $\Delta = 4$ in Eq.~(\ref{spinham}). It is now 
interesting to also consider the more general case of tunable 
anisotropy, which could for example be realized with dimers consisting 
of two different atomic species \cite{MVDPeHub}. As the hopping becomes
stronger, the perturbative analysis used in the previous sections 
becomes unreliable.  It is known that the model of Eq.~(\ref{eq:xxzHam})
is critical for $-1 \leq \Delta \leq 1$. Therefore, the perturbation 
treatment breaks down exactly at the point where the hopping becomes 
equal to or larger than the nearest neighbor dimer-dimer interaction. 
However, it is still possible to use the field theoretical methods 
of Sec.~\ref{sec:ftapr} to calculate the correlation functions and 
expectation values.

In the critical region excitations are gapless in the thermodynamic limit, 
so that there is no incompressible phase at half-filling. The crossover 
between the completely filled and completely empty regions in 
Fig.~\ref{fig:AF_trap} is therefore continuous as a function of the
effective field and there is no extended half-filled phase. 
The strength of hopping is therefore crucial for the behavior of the system:
weak hopping enables the presence of a compressible phase between the 
incompressible ferromagnetic and antiferromagnetic phases. 
With increasing the hopping strength the incompressible antiferromagnetic 
phase shrinks and completely vanishes when the hopping reaches 
the value of the nearest neighbor interaction, $\Delta \leq 1$.

\begin{figure}[t]
\centerline{\includegraphics[width=0.48\textwidth]{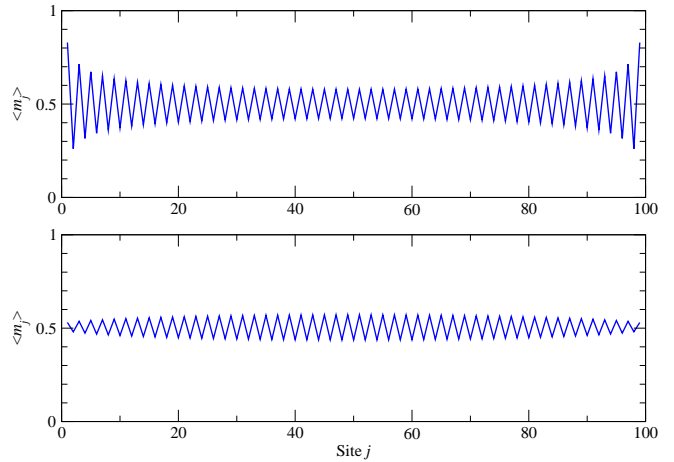}}
\caption{Top: Ground state density profile in a homogeneous lattice 
with $N_s = 99$ sites and hard-wall boundaries, for $N=50$ particles,
but with hopping strength equal to the nearest neighbor repulsion, 
$\Delta=1$ (compare to Fig.~\ref{fig:density}, top panel).
Bottom: The same, but without the effective edge fields in 
the spin chain model.}
\label{fig:density2}
\end{figure}

Equally interesting is the effect of hopping on the density of 
dimers along the chain.  As discussed in Sec.~\ref{sec:propcp}, 
density-wave modulations appear in Fig.~\ref{fig:density} because of 
the effective motion of kinks. The complex interplay between the
kinetic and interaction terms in the critical region now leads to 
a further modification of the density pattern along the chain 
\cite{neel,prl95,rommer}. In particular, the amplitude of
the ground state density oscillations is now significantly reduced 
towards the middle of the chain with a characteristic drop-off as 
shown in Fig.~\ref{fig:density2}. Excited states with larger $N$ 
would then exhibit modulations on top of this ground state 
pattern, similar to Fig.~\ref{fig:density}.

Interestingly, the exact form of the boundary conditions play now 
a much more important role. Namely, the effective edge field 
of the spin chain model discussed in Sec~\ref{sec:kink} accounts 
for a large part of the ground state density oscillations. For 
comparison, in Fig.~\ref{fig:density2} we also show the density for 
the spin chain model without any edge field. The density amplitude 
is now smaller near the boundary. At $\Delta=1$, this amplitude 
has been predicted to follow approximately a $\sqrt{\sin (\pi x/N_s)}$
behavior \cite{neel}, which however is strongly affected by temperature
\cite{prl95,rommer} due to the gapless modes.

\section{Summary}

In this paper, we have studied the many-body dynamics of 
attractively bound pairs of atoms in the Bose-Hubbard model. 
When the on-site interaction between the atoms exceeds by a 
sufficient amount the bandwidth of the lowest single-particle 
Bloch band, the pairs are well co-localized and can be treated 
as composite dimer particles. Then the effective model for dimers 
on a lattice is equivalent to an asymmetric spin-$\frac{1}{2}$ 
$XXZ$ model in an external magnetic field: the nearest neighbor 
interaction between the dimers translates into an Ising-type 
spin-spin interaction and the dimer tunneling to a spin-spin 
coupling in the $x-y$ plane. 

The case of repulsively bound pairs studied in \cite{PSAF} corresponds 
to a ferromagnetic Ising coupling. In contrast, for attractively bound 
pairs analyzed here, the Ising coupling is antiferromagnetic leading 
to a much richer phase diagram. The asymmetry parameter $\Delta$
of the $XXZ$ model is equal to 4; as a result, the system is gapped 
for both zero and full magnetization. The zero magnetization
state, corresponding to exactly half filling of dimers, exhibits
antiferromagnetic order, while the full magnetization (ferromagnetic)
states correspond to vanishing or full filling of dimers. 
When the effective magnetic field, or for that matter the chemical 
potential for the dimers, exceeds critical values, the gap is 
closed and the dimer system becomes critical with finite compressibility. 

We have derived the critical values of the chemical potential for 
the transition points from the gapped ferromagnetic and antiferromagnetic
phases to the compressible phases in 1D, employing the known exact 
solutions of the $XXZ$ model, and by using a strong-coupling approximation
in higher dimensions. The properties of the compressible phases
are quite different in one and higher spatial dimensions. 
In 1D, close to half filling, the system can be well described by 
kink-like domain walls which separate antiferromagnetic strings
of opposite phase and can propagate through the lattice almost freely.
In higher dimensions, the motion of similar defects is strongly suppressed.
A simple approximate description in terms of non-interacting kinks gives
rather accurate predictions for the dimer density as well as non-local 
density-density correlations. 

In order to explain the first order correlations, we employed a field 
theoretical approach based on the Luttinger liquid theory. The corresponding
Luttinger parameter was obtained by solving the Bethe Ansatz equations 
for the equivalent $XXZ$ model in the regime of critical magnetic fields.
The expressions for the first-order and density-density correlations 
showed remarkably good agreement with the numerical data obtained by 
DMRG simulations. Finally we discussed the consequences of changing 
the anisotropy parameter of the $XXZ$ model. Our studies attest that 
interaction bound pairs of atoms in deep optical lattices can provide 
a versatile tool to simulate and explore quantum spin models.

\begin{acknowledgments}
We are thankful for discussions with Fabian Essler, Manuel Valiente and Xue-Feng Zhang.
This work was supported by the EU network EMALI and the DFG
through the SFB TR49.
\end{acknowledgments}



\end{document}